\documentclass[aps,prl,twocolumn,tightenlines,superscriptaddress,showpacs,byrevtex]{revtex4}

\usepackage{rotating}
\usepackage{longtable}
\usepackage{amsmath}
\usepackage{graphicx} % Include figure files
\usepackage{dcolumn}  % Align table columns on decimal point
\usepackage{multirow}
\usepackage{color}

\renewcommand{\arraystretch}{1.1}

\newcommand{\mev}{\mathrm{MeV}}

\newcommand{\mevm}{\mathrm{MeV}/c^2}

\newcommand{\gevm}{\mathrm{GeV}/c^2}

\newcommand{\ee}{e^+e^-}

\newcommand{\pp}{\pi^+\pi^-}
\newcommand{\ga}{\gamma}

\newcommand{\Uf}{\Upsilon(5S)}
\newcommand{\Uo}{\Upsilon(1S)}
\newcommand{\Ud}{\Upsilon(1D)}
\newcommand{\Ut}{\Upsilon(2S)}
\newcommand{\Uth}{\Upsilon(3S)}
\newcommand{\Un}{\Upsilon(nS)}

\newcommand{\mm}{M_{\rm miss}}
\newcommand{\mmpp}{M_{\rm miss}(\pi^+\pi^-)}

\newcommand{\et}{\eta_b(1S)}
\newcommand{\ett}{\eta_b(2S)}
\newcommand{\etn}{\eta_b(nS)}
\newcommand{\etm}{\eta_b(mS)}
\newcommand{\hb}{h_b(1P)}
\newcommand{\hbp}{h_b(2P)}
\newcommand{\hbn}{h_b(nP)}

\newcommand{\ks}{K^0_S}

\newcommand{\km}{K^-}
\newcommand{\pip}{\pi^{+}}
\newcommand{\pipm}{\pi^{\pm}}

\newcommand{\fb}{\mathrm{fb}^{-1}}
\newcommand{\br}{\mathcal{B}}

\newcommand{\etal}{\em et al.}

\newcommand{\dmmppg}{M^{(n)}_{\rm miss}(\pi^+\pi^-\gamma)}
\newcommand{\dmo}{M^{(1)}_{\rm miss}(\pi^+\pi^-\gamma)}
\newcommand{\dmt}{M^{(2)}_{\rm miss}(\pi^+\pi^-\gamma)}
\newcommand{\zb}{Z_b}
\newcommand{\zbo}{Z_b(10610)}
\newcommand{\zbt}{Z_b(10650)}

\begin{document}

\title{\boldmath Evidence for the $\eta_b(2S)$ and observation of
  $h_b(1P)\to\et\gamma$ and $h_b(2P)\to\et\gamma$}
\date{20 July 2012}
\begin{abstract}
\noindent
We report the first evidence for the $\ett$ using the $\hbp\to\ett\ga$
transition and the first observation of the $\hb\to\et\ga$ and
$\hbp\to\et\ga$ transitions. The mass and width of the $\et$ and
$\ett$ are measured to be $m_{\et}=(9402.4\pm1.5\pm1.8)\,\mevm$,
$m_{\ett}=(9999.0\pm3.5\,^{+2.8}_{-1.9})\,\mevm$ and
$\Gamma_{\et}=(10.8\,^{+4.0}_{-3.7}\,^{+4.5}_{-2.0})\,\mev$.
We also update the $\hb$ and $\hbp$ mass measurements.
We use a $133.4\,{\rm fb}^{-1}$ data sample collected at energies near
the $\Uf$ resonance with the Belle detector at the KEKB
asymmetric-energy $\ee$ collider.
\end{abstract}


%%% Paper:    eta_b(2S), eta_b(1S)
%%% Journal:  Physical Review Letters
%%% Contacts: R. Mizuk (mizuk@itep.ru)
%%%           A. Bondar (bondar@inp.nsk.su)
%%% Non-responding authors or those who said NO are commented out.
%%% ====================================================================
%%% Click the RELOAD button on your web browser to see the updated file.
%%% ====================================================================
%%% Use \input{author} to insert this material into your latex file.
%%%%% Force institutions to appear in alphabetical order when typeset.
%%%\affiliation{University of Bonn, Bonn}
\affiliation{Budker Institute of Nuclear Physics SB RAS and Novosibirsk State University, Novosibirsk 630090}
\affiliation{Faculty of Mathematics and Physics, Charles University, Prague}
%%%\affiliation{Chiba University, Chiba}
\affiliation{University of Cincinnati, Cincinnati, Ohio 45221}
%%%\affiliation{Department of Physics, Fu Jen Catholic University, Taipei}
\affiliation{Justus-Liebig-Universit\"at Gie\ss{}en, Gie\ss{}en}
\affiliation{Gifu University, Gifu}
%%%\affiliation{The Graduate University for Advanced Studies, Hayama}
\affiliation{Gyeongsang National University, Chinju}
\affiliation{Hanyang University, Seoul}
\affiliation{University of Hawaii, Honolulu, Hawaii 96822}
\affiliation{High Energy Accelerator Research Organization (KEK), Tsukuba}
%%%\affiliation{Hiroshima Institute of Technology, Hiroshima}
%%%\affiliation{University of Illinois at Urbana-Champaign, Urbana, Illinois 61801}
\affiliation{Indian Institute of Technology Guwahati, Guwahati}
\affiliation{Indian Institute of Technology Madras, Madras}
%%%\affiliation{Indiana University, Bloomington, Indiana 47408}
\affiliation{Institute of High Energy Physics, Chinese Academy of Sciences, Beijing}
\affiliation{Institute of High Energy Physics, Vienna}
\affiliation{Institute of High Energy Physics, Protvino}
%%%\affiliation{Institute of Mathematical Sciences, Chennai}
\affiliation{INFN - Sezione di Torino, Torino}
\affiliation{Institute for Theoretical and Experimental Physics, Moscow}
\affiliation{J. Stefan Institute, Ljubljana}
\affiliation{Kanagawa University, Yokohama}
\affiliation{Institut f\"ur Experimentelle Kernphysik, Karlsruher Institut f\"ur Technologie, Karlsruhe}
\affiliation{Korea Institute of Science and Technology Information, Daejeon}
\affiliation{Korea University, Seoul}
%%%\affiliation{Kyoto University, Kyoto}
\affiliation{Kyungpook National University, Taegu}
\affiliation{\'Ecole Polytechnique F\'ed\'erale de Lausanne (EPFL), Lausanne}
\affiliation{Faculty of Mathematics and Physics, University of Ljubljana, Ljubljana}
\affiliation{Luther College, Decorah, Iowa 52101}
\affiliation{University of Maribor, Maribor}
\affiliation{Max-Planck-Institut f\"ur Physik, M\"unchen}
\affiliation{University of Melbourne, School of Physics, Victoria 3010}
\affiliation{Graduate School of Science, Nagoya University, Nagoya}
\affiliation{Kobayashi-Maskawa Institute, Nagoya University, Nagoya}
%%%\affiliation{Nara University of Education, Nara}
\affiliation{Nara Women's University, Nara}
\affiliation{National Central University, Chung-li}
\affiliation{National United University, Miao Li}
\affiliation{Department of Physics, National Taiwan University, Taipei}
\affiliation{H. Niewodniczanski Institute of Nuclear Physics, Krakow}
\affiliation{Nippon Dental University, Niigata}
\affiliation{Niigata University, Niigata}
\affiliation{University of Nova Gorica, Nova Gorica}
\affiliation{Osaka City University, Osaka}
%%%\affiliation{Osaka University, Osaka}
\affiliation{Pacific Northwest National Laboratory, Richland, Washington 99352}
%%%\affiliation{Panjab University, Chandigarh}
%%%\affiliation{Peking University, Beijing}
%%%\affiliation{Princeton University, Princeton, New Jersey 08544}
\affiliation{Research Center for Electron Photon Science, Tohoku University, Sendai}
%%%\affiliation{Research Center for Nuclear Physics, Osaka University, Osaka}
%%%\affiliation{RIKEN BNL Research Center, Upton, New York 11973}
%%%\affiliation{Saga University, Saga}
\affiliation{University of Science and Technology of China, Hefei}
\affiliation{Seoul National University, Seoul}
%%%\affiliation{Shinshu University, Nagano}
\affiliation{Sungkyunkwan University, Suwon}
\affiliation{School of Physics, University of Sydney, NSW 2006}
\affiliation{Tata Institute of Fundamental Research, Mumbai}
\affiliation{Excellence Cluster Universe, Technische Universit\"at M\"unchen, Garching}
%%%\affiliation{Toho University, Funabashi}
\affiliation{Tohoku Gakuin University, Tagajo}
\affiliation{Tohoku University, Sendai}
\affiliation{Department of Physics, University of Tokyo, Tokyo}
\affiliation{Tokyo Institute of Technology, Tokyo}
\affiliation{Tokyo Metropolitan University, Tokyo}
\affiliation{Tokyo University of Agriculture and Technology, Tokyo}
%%%\affiliation{Toyama National College of Maritime Technology, Toyama}
\affiliation{CNP, Virginia Polytechnic Institute and State University, Blacksburg, Virginia 24061}
\affiliation{Wayne State University, Detroit, Michigan 48202}
\affiliation{Yamagata University, Yamagata}
\affiliation{Yonsei University, Seoul}
  \author{R.~Mizuk}\affiliation{Institute for Theoretical and Experimental Physics, Moscow} % ITEP
  \author{D.~M.~Asner}\affiliation{Pacific Northwest National Laboratory, Richland, Washington 99352} % PNNL
  \author{A.~Bondar}\affiliation{Budker Institute of Nuclear Physics SB RAS and Novosibirsk State University, Novosibirsk 630090} % BINP
  \author{T.~K.~Pedlar}\affiliation{Luther College, Decorah, Iowa 52101} % Luther
  \author{I.~Adachi}\affiliation{High Energy Accelerator Research Organization (KEK), Tsukuba} % KEK
% \author{K.~Adamczyk}\affiliation{H. Niewodniczanski Institute of Nuclear Physics, Krakow} % Krakow
  \author{H.~Aihara}\affiliation{Department of Physics, University of Tokyo, Tokyo} % Tokyo
  \author{K.~Arinstein}\affiliation{Budker Institute of Nuclear Physics SB RAS and Novosibirsk State University, Novosibirsk 630090} % BINP
% \author{Y.~Arita}\affiliation{Graduate School of Science, Nagoya University, Nagoya} % Nagoya
% \author{T.~Aso}\affiliation{Toyama National College of Maritime Technology, Toyama} % Toyama
  \author{V.~Aulchenko}\affiliation{Budker Institute of Nuclear Physics SB RAS and Novosibirsk State University, Novosibirsk 630090} % BINP
  \author{T.~Aushev}\affiliation{Institute for Theoretical and Experimental Physics, Moscow} % ITEP
  \author{T.~Aziz}\affiliation{Tata Institute of Fundamental Research, Mumbai} % Tata
% ...
% BINP
% \author{P.~Zhou}\affiliation{Wayne State University, Detroit, Michigan 48202} % WayneState
% \author{V.~Zhulanov}\affiliation{Budker Institute of Nuclear Physics SB RAS and Novosibirsk State University, Novosibirsk 630090} % BINP
% \author{T.~Zivko}\affiliation{J. Stefan Institute, Ljubljana} % Ljubljana
% \author{A.~Zupanc}\affiliation{Institut f\"ur Experimentelle Kernphysik, Karlsruher Institut f\"ur Technologie, Karlsruhe} % Karlsruhe
% \author{N.~Zwahlen}\affiliation{\'Ecole Polytechnique F\'ed\'erale de Lausanne (EPFL), Lausanne} % Lausanne
% \author{O.~Zyukova}\affiliation{Budker Institute of Nuclear Physics SB RAS and Novosibirsk State University, Novosibirsk 630090} % BINP
\collaboration{The Belle Collaboration}

\maketitle

{\renewcommand{\thefootnote}{\fnsymbol{footnote}}}
\setcounter{footnote}{0}

Bottomonium, a bound system of a $b \bar b$ quark-antiquark pair, is
well described by nonrelativistic quantum mechanics due to the slow
motion of the heavy quarks~\cite{nora}. The spin-singlet ($S=0$)
states of the quark pair with zero orbital momentum, $L=0$, are
customarily called $\etn$ with $n=1,2, \ldots$. The hyperfine
splitting $\Delta M_{\rm HF}$ from the corresponding spin-triplet
state, {\it i.e.} the mass difference between the $\Un$ and $\etn$,
provides an important measure of the spin-spin interaction between the
quark and the antiquark. The existing
measurements~\cite{BaBar_CLEO_etab} of the $\et$ mass are only in
marginal agreement with theoretical expectations~\cite{pQCD,Lattice} 
and the $\et$ width is yet to be measured. There is no information
available on the radially excited state $\ett$.

In this Letter, we report the first evidence for the $\ett$ in the
$\hbp\to\ett\gamma$ transition and the first observation of the
$\hb\to\et\gamma$ and $\hbp\to\et\gamma$ transitions.
We use a $121.4\,\fb$ data sample at the $\Uf$ resonance and
$12.0\,\fb$ of energy-scan data collected nearby with the Belle
detector~\cite{BELLE_DETECTOR} at the KEKB asymmetric-energy $\ee$
collider~\cite{KEKB}.

We study the processes $\ee\to\Uf\to\hbn\pp\to[\etm\ga]\pp$ in which
the $\etm$ states are reconstructed inclusively. The approximate
values of the expected energies of the photons in the $\hbn$
rest frame are given in Table~\ref{tab:pi0_veto}. The $\hbn$ signal is
tagged using the missing mass of the $\pp$ pair $\mmpp$, while the
$\etm$ signal is tagged using the variable
$\dmmppg\equiv\mm(\pp\ga)-\mmpp+m_{\hbn}$. The missing mass is defined
via
$\mm(X)=\sqrt{(E_{\rm c.m.}-E_X^*)^2-p_X^{*2}}$,
where $E_{\rm c.m.}$ is the center-of-mass (c.m.) energy and $E^*_X$
and $p^*_X$ are the energy and momentum of the system $X$ measured in
the c.m.\ frame.
We fit the $\mmpp$ spectra for different $\dmmppg$ bins to measure the
$\hbn$ yield. This procedure removes the background due to random
$\pp$ combinations. The $\hbn$ yield peaks at $\dmmppg$ values
corresponding to the masses of the $\etm$ states from the
$\hbn\to\etm\ga$ transitions.

\begin{table}[htb]
\caption{Expected photon energies in $\hbn\to\etm\ga$ transitions
  ($E_\gamma$), $\pi^0$ veto parameters ($\Delta M$ and $E_{\rm th}$)
  and the parameter $\sigma$ of the $\dmmppg$ resolution function.}
\label{tab:pi0_veto}
\renewcommand{\arraystretch}{1.1}
\begin{ruledtabular}
\begin{tabular}{l|cccc}
& $E_{\gamma}\,,$ & $\Delta M,$ & $E_{\rm th}\,,$ & $\sigma,$ \\
& $\mev$         & $\mevm$     & $\mev$         & $\mevm$ \\
\hline
$\hbp\to\ett\gamma$ & 260 & 10 & 125 & $13.6\pm1.1$ \\
$\hb\to\et\gamma$   & 500 & 13 & 75  & $19.8\pm1.1$ \\
$\hbp\to\et\gamma$  & 860 & 17 & 75  &  $8.6\pm0.7$
\end{tabular}
\end{ruledtabular}
\end{table}

The selection criteria for the $\pp$ pairs are the same as those
described in Ref.~\cite{Belle_hb}. We use events that pass the
Belle-standard hadronic event selection and consider all positively
identified $\pp$ pairs that originate from the vicinity of the
interaction point. Belle previously observed that the decay
$\Uf\to\hbn\pp$ proceeds via the intermediate resonances $\zbo$ and
$\zbt$~\cite{Belle_zb}. We exploit this with the additional
requirement $10.59\,\gevm<\mm(\pipm)<10.67\,\gevm$, which suppresses
the combinatorial background by a factor of 5 [1.6] for the $\hb$
[$\hbp$] without any significant loss of the signal.
Photon candidates are clusters in the electromagnetic calorimeter that
are not associated with charged tracks. We apply a veto on
$\pi^0\to\ga\ga$ decays, rejecting a photon candidate if the invariant
mass of it and any other photon in the event with energy above the
threshold $E_{\rm th}$ is within $\Delta M$ of the $\pi^0$ mass. The
parameters $E_{\rm th}$ and $\Delta M$, listed in
Table~\ref{tab:pi0_veto}, are chosen by maximizing the ratio
$\frac{S}{\sqrt{B}}$, where $S$ is the number of signal events in the
Monte Carlo (MC) simulation and $B$ is the number of background events
estimated from a small fraction (0.1\%) of the data.
To suppress continuum $\ee\to q\bar{q}$ ($q=u$, $d$, $s$, $c$)
background, we use the ratio $R_2$ of the second- to zeroth-order
Fox-Wolfram moments~\cite{Fox-Wolfram}. In the $\et$ analysis, we
require $R_2<0.3$, which was optimized using the $\Uf\to\Ut\pp$
decays~\cite{Belle_hb}. For the high statistics $\hb\to\et\ga$
transition, the optimum is shifted to $R_2<0.32$, which we adopt here
for the $\ett$ analysis.

To calibrate the photon energy resolution function, we use three
control channels: $D^{*0}\to\ga D^0(\to\km\pip)$, $\pi^0\to\ga\ga$ and
$\eta\to\ga\ga$. For the two-photon final states, we require that the
energies of the photons in the laboratory frame be almost equal:
$|E_1-E_2|/(E_1+E_2)<0.05$. This reduces the resolution shape
dependence to a single variable. The resolution shape is parameterized
by a double-sided bifurcated Crystal Ball function~\cite{CB} in which
a bifurcated Gaussian is smoothly joined with power law tails on both
sides.
The signal is extracted using the
$M(K^-\pi^+\gamma)-M(K^-\pi^+)+m_{D^0}$ distribution for the $D^{*0}$
and $M(\gamma\gamma)$ for the $\pi^0$ and $\eta$. From comparisons of
the peak positions and widths in data and MC simulation, we determine
shifts in the photon energy $\Delta E/E$ and width correction factors
$f$ as a function of $E$. Various calibration channels give consistent
results; an uncertainty is assigned based on their spread. These
translate to typical mass shift and width-correction factors of
$1.2\,\mevm$ and $1.13$, respectively. Average values of left and
right widths ($\sigma$) of the bifurcated Gaussian components of the
$\dmmppg$ resolution functions are given in Table~\ref{tab:pi0_veto}.

\begin{figure}[htbp]
\includegraphics[width=7cm]{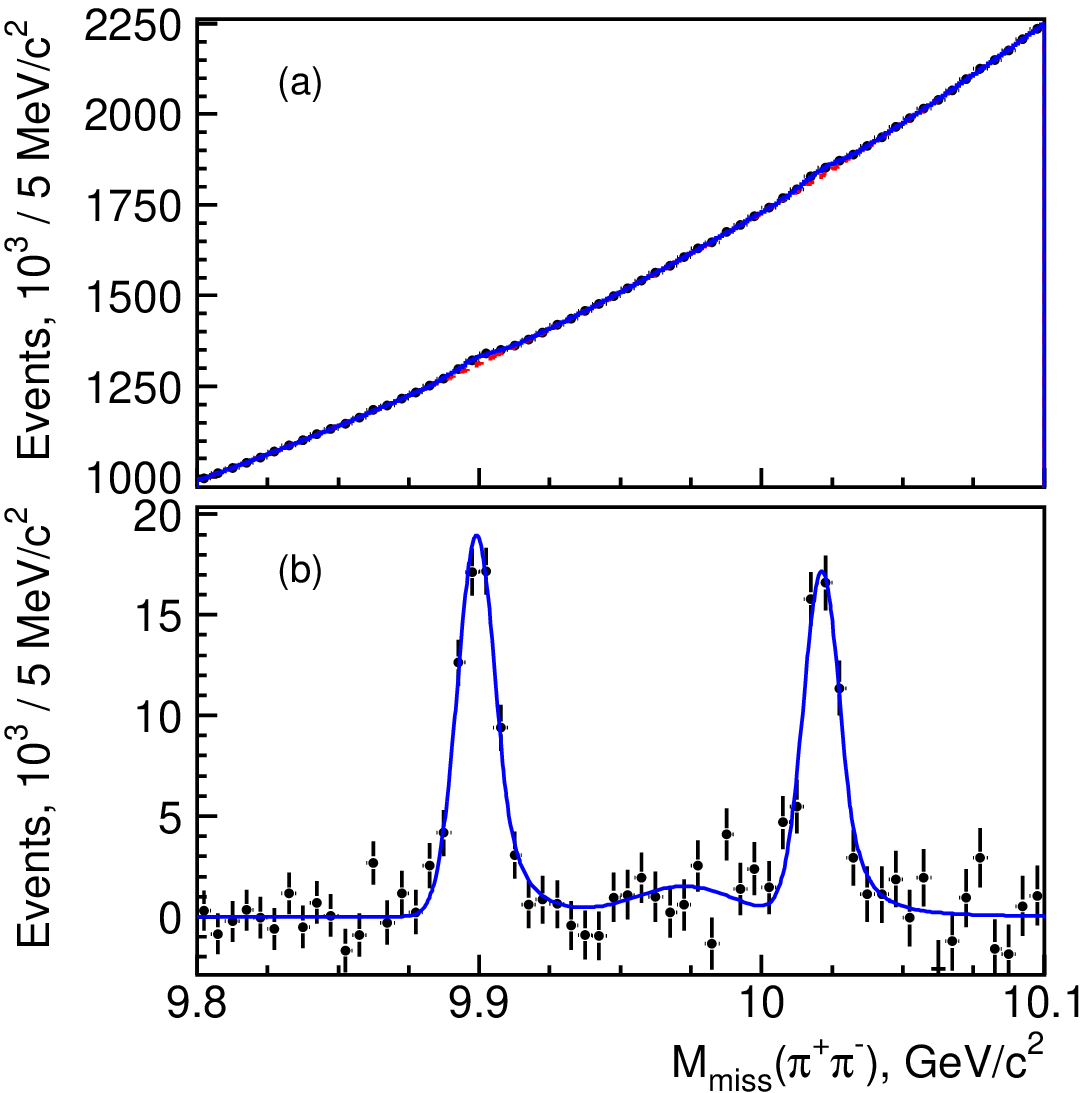}
\caption{(colored online) The $\mmpp$ spectrum in the
  $\hb$ region. In (a) the data are the points with error bars with
  the fit function (blue solid curve) and background (red dashed
  curve) overlaid. (b) shows the background subtracted data (points
  with error bars) while the signal component of the fit is overlaid
  (blue curve). The background is combinatorial.}
\label{mmpp_hb}
\end{figure}

\begin{figure}[htbp]
\includegraphics[width=7cm]{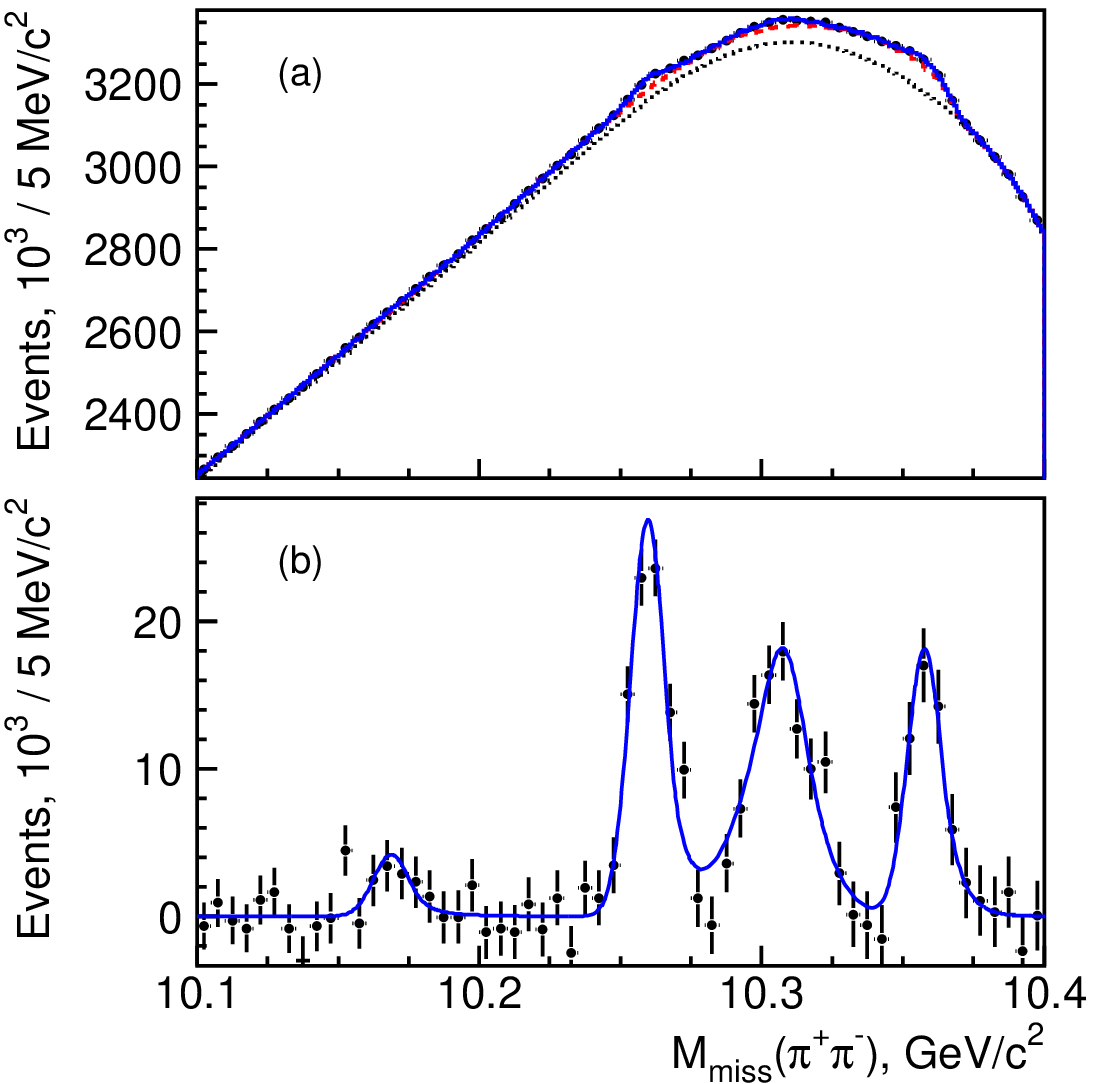}
\caption{(colored online) The $\mmpp$ spectrum in the
  $\hbp$ region. The legend is the same as in Fig.~\ref{mmpp_hb}. The
  background components are random combinations and $\ks\to\pp$
  decays. The combinatorial component is shown in (a) by the black
  dotted curve.}
\label{mmpp_hb2}
\end{figure}

We update previous Belle measurements of the $\hbn$
masses~\cite{Belle_hb}, incorporating a 10\% increase in statistics
and the requirement of an intermediate $\zb$~\cite{mmpp_fit_details}.
The results of the fits are shown in Figs~\ref{mmpp_hb} and
\ref{mmpp_hb2} and the fitted signal parameters are listed in
Table~\ref{tab:mmpp_fits}.
\begin{table}[tb!h]
\caption{The yield and mass of peaking components from the fits to the
  $\mmpp$. Here and everywhere in this Letter, the first quoted
  uncertainty is statistical (unless stated otherwise) and the
  second (if present) is systematic.}
\label{tab:mmpp_fits}
\renewcommand{\arraystretch}{1.1}
\begin{ruledtabular}
\begin{tabular}{c|cc}
& $N,\;10^3$ & Mass, $\mevm$ \\
\hline
$\Uf\to\hb$  & $70.3\pm3.3^{+1.9}_{-0.7}$ & $9899.1\pm0.4\pm1.0$ \\
$\Uth\to\Uo$ & $13\pm7$                  & $9973.0$ \\
$\Uf\to\Ut$  & $61.3\pm4.1$              & $10021.3\pm0.5$ \\
\hline
$\Uf\to\Ud$  & $14\pm7$                  & $10169\pm3$ \\
$\Uf\to\hbp$ & $89.5\pm6.1^{+0.0}_{-5.8}$ & $10259.8\pm0.5\pm1.1$ \\
$\Ut\to\Uo$  & $97\pm12$                 & $10305.6\pm1.2$ \\
$\Uf\to\Uth$ & $58\pm8$                  & $10357.7\pm1.0$ \\
\end{tabular}
\end{ruledtabular}
\end{table}
The confidence level of the fit in the $\hb$ [$\hbp$] region is 35\%
[70\%].
The estimation of systematic uncertainties follows
Ref.~\cite{Belle_hb}.
For hyperfine splittings
$\sum\limits^2_{J=0}\frac{2J+1}{9}m_{\chi_{bJ}(nP)}-m_{\hbn}$, we find
$\Delta M_{\rm HF}(1P)=(+0.8\pm1.1)\,\mevm$ and $\Delta M_{\rm
  HF}(2P)=(+0.5\pm1.2)\,\mevm$, where statistical and systematic
uncertainties in mass are added in quadrature.

\begin{figure}[tbh]
\includegraphics[width=7cm]{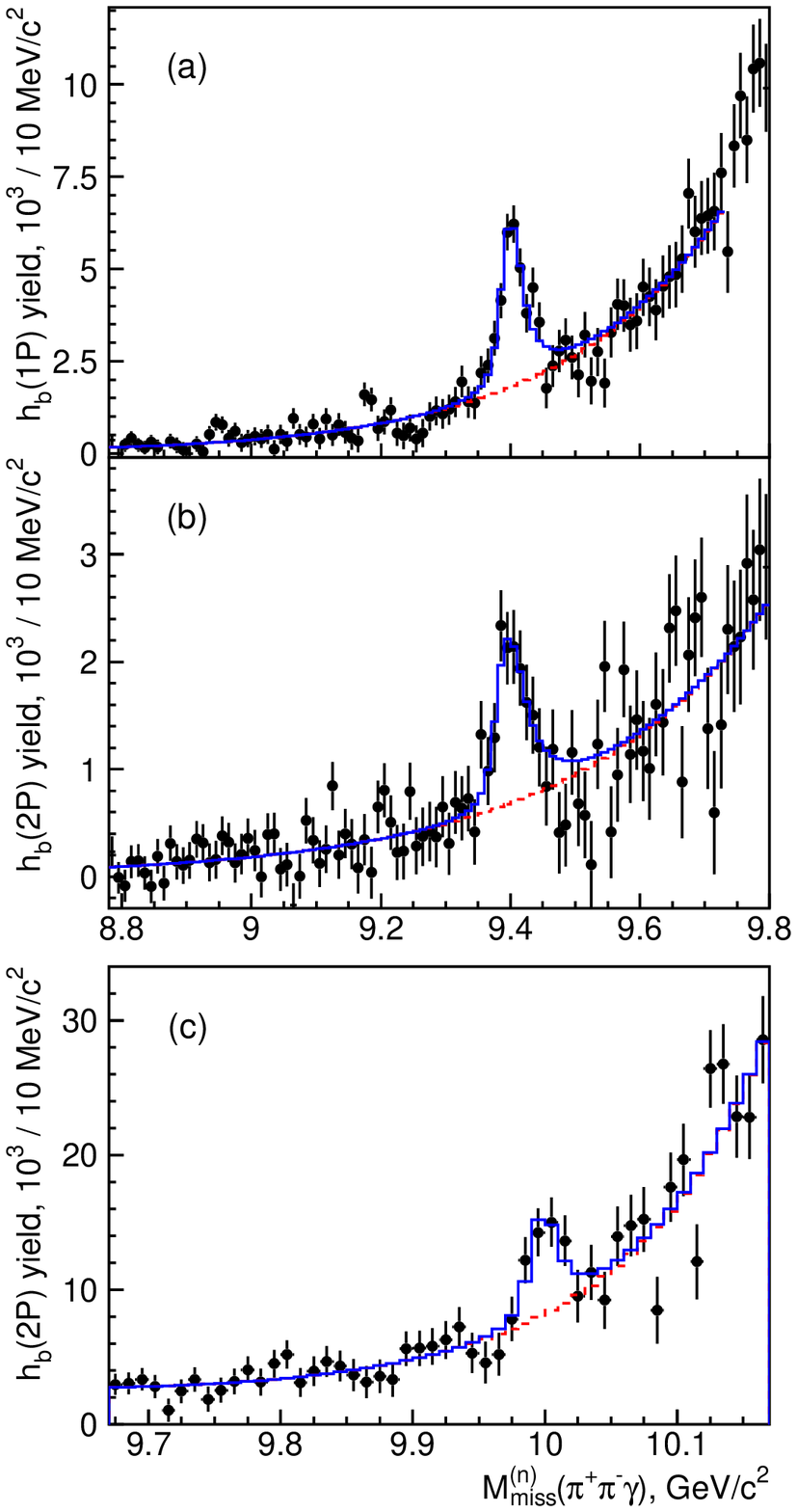}
\caption{The $\hb$ yield {\it vs.} $\dmo$ (a), and $\hbp$ yield {\it
    vs.}  $\dmt$ in the $\et$ region (b) and in the $\ett$ region
  (c). The solid (dashed) histogram is the fit result (background
  component of the fit function).}
\label{hbn_vs_dmmppg}
\end{figure}

We fit the $\mmpp$ spectra for each $\dmmppg$ bin to measure the
$\hbn$ yield as a function of $\dmmppg$. We fix the masses of the
peaking components at the values given in Table~\ref{tab:mmpp_fits}.
Although the $\mmpp$ combinatorial background shape in the $\hbp$
region is rather complicated (described by an eighth-order Chebyshev
polynomial), it changes slowly with $\dmt$. For the $\mmpp$ fits for
each bin, we multiply the polynomial shape with parameters fixed at
their values from the overall fit by a lower-order polynomial with
floating coefficients; this procedure improves the accuracy of the
$\hbn$ yield measurements. For this lower-order function, we use a
first- [fifth-] order polynomial for the $\et$ [$\ett$] $\dmt$ region.
From a generic MC simulation, we find that the $\ks\to\pp$
contribution is independent of the $\dmt$ value in the $\et$ region;
in the $\ett$ region, we restrict the $\mmpp$ fit range to
$10.10\,\gevm-10.34\,\gevm$, thereby avoiding the sharp rise in the
$\ks\to\pp$ contribution that occurs at $10.37\,\gevm$.
The results for the $\hb$ and $\hbp$ yields as a function of 
$\dmmppg$ are presented in Fig.~\ref{hbn_vs_dmmppg}. 
Clear peaks at $9.4\,\gevm$ and $10.0\,\gevm$ are identified as
signals for the $\et$ and $\ett$, respectively. Generic MC simulations
indicate that no peaking backgrounds are expected in these spectra.

We fit the $\hbn$ yield dependence on $\dmmppg$ to a sum of the $\etn$
signal components described by the convolution of a non-relativistic
Breit-Wigner function with the calibrated resolution function
described above and a background parameterized by an exponentiation of
a first- [second-] order polynomial in the $\et$ [$\ett$] region. The
two $\dmmppg$ spectra [from the $\hb$ and $\hbp$] with $\et$ signals
are fitted simultaneously.
We take into account the effect of multiple photons in the same
$\dmmppg$ bin by increasing the errors in the corresponding $\mmpp$
histogram.
We find event yields for the $\hbn\to\etm$ transitions of
$N_{1P \to1S}=(23.5\pm2.0)\times10^3$,
$N_{2P \to1S}=(10.3\pm1.3)\times10^3$ and
$N_{2P \to2S}=(25.8\pm4.9)\times10^3$;
the fitted masses and width are
$m_{\et}=(9402.4\pm1.5\pm1.8)\,\mevm$,
$\Gamma_{\et}=(10.8\,^{+4.0}_{-3.7}\,^{+4.5}_{-2.0})\,\mev$
and 
$m_{\ett}=(9999.0\pm3.5\,^{+2.8}_{-1.9})\,\mevm$.
The confidence level of the $\et$ [$\ett$] fit is 61\% [36\%].
If the $\ett$ width is allowed to float in the fit, we find
$\Gamma_{\ett}=(4^{+12}_{-20})\,\mev$ or $\Gamma_{\ett}<24\,\mev$ at
90\%~C.L. using the Feldman-Cousins approach~\cite{feld_cous}. For the
$\ett$ mass determination and yield measurements quoted above, we fix
the $\ett$ width at its value from perturbative
calculations~\cite{ett_wid}
$\Gamma_{\ett}=\Gamma_{\et}\frac{\Gamma_{ee}^{\Ut}}{\Gamma_{ee}^{\Uo}}
=(4.9^{+2.7}_{-1.9})\,\mev$, where the uncertainty is due to our
experimental uncertainty in $\Gamma_{\et}$.

To estimate the systematic uncertainties in the $\etn$ parameters, we
vary the polynomial orders and fit intervals in the $\mmpp$ \&
$\dmmppg$ fits, and the $\dmmppg$ binning by scanning the starting
point of the $10\,\mevm$ bin with $1\,\mevm$ steps. We also multiply
the non-relativistic Breit-Wigner function by an $E_{\gamma}^3$ term
expected for an electric dipole transition and include the uncertainty
in the $\hb$ and $\hbp$ masses and in the estimated value of the
$\ett$ width. The contribution of each source is given in
Table~\ref{tab:syst_etn_par}.
We add the various contributions in quadrature to estimate the total
systematic uncertainty.
We repeat the analysis using signal MC simulation instead of data and
find no shift of the $\etn$ parameters compared to the MC input.
For the hyperfine splittings $m_{\Un}-m_{\etn}$ we determine $\Delta
M_{\rm HF}(1S)=(57.9\pm2.3)\,\mevm$ and
$\Delta M_{\rm HF}(2S)=(24.3^{+4.0}_{-4.5})\,\mevm$, where statistical and
systematic uncertainties in mass are added in quadrature.
\begin{table}[tbh]
\caption{ Systematic uncertainties in the $\et$ and $\ett$ masses (in
  $\mevm$) and in the $\et$ width (in $\mev$).}
\label{tab:syst_etn_par}
\renewcommand{\arraystretch}{1.4}
\begin{ruledtabular}
\begin{tabular}{l|ccc}
& $m_{\et}$ & $\Gamma_{\et}$ & $m_{\ett}$ \\
\hline
$\mmpp$ fits      & $^{+0.1}_{-0.2}$ & $^{+0.0}_{-0.8}$ & $^{+0.0}_{-0.1}$ \\
$\dmmppg$ binning & $^{+0.2}_{-0.5}$ & $^{+0.7}_{-0.8}$ & $^{+2.3}_{-1.4}$ \\
$\dmmppg$ fit     & $^{+0.2}_{-0.0}$ & $^{+4.2}_{-0.6}$ & $^{+0.5}_{-0.2}$ \\
Calibration       & $\pm1.4$        & $\pm1.5$        & $\pm0.7$ \\
$\etn$ line shape & $^{+0.6}_{-0.0}$ & $^{+0.0}_{-0.4}$ & $^{+0.6}_{-0.0}$ \\
$\hbn$ mass       & $\pm0.9$        & $\pm0.0$        & $\pm1.1$ \\
$\ett$ width      & -               & -               & $^{+0.1}_{-0.0}$ \\
\hline 
Total             & $\pm1.8$        & $^{+4.5}_{-2.0}$ & $^{+2.8}_{-1.9}$ \\
\end{tabular}
\end{ruledtabular}
\end{table}

Using Wilks' theorem~\cite{wilks}, we find $15\,\sigma$ [$9\,\sigma$]
for the $\hb\to\et\ga$ [$\hbp\to\et\ga$] statistical significance.
For the significance of the $\ett$ signal, we use a method that takes
into account the trial factor associated with the definition of the
search window or the so-called ``look-elsewhere
effect''~\cite{look_else}. To determine this window, we conservatively
assume $r=0$ and $r=1$ for the ratio 
$r=\frac{\Delta M_{\rm HF}(2S)}{\Delta M_{\rm HF}(1S)}$.  (For
reference, the measured value of $r = 0.420^{+0.071}_{-0.079}$ is
consistent with perturbative calculations~\cite{dmhf_rat} that predict
$\frac{m_{\Ut}^2}{m_{\Uo}^2}\frac{\Gamma_{ee}^{\Ut}}{\Gamma_{ee}^{\Uo}}
=0.513\pm0.011$, where the error is due to the uncertainties in
$\Gamma_{ee}$.)
We find the significance of the $\ett$ signal to be $4.8\,\sigma$
($4.2\,\sigma$ including systematics).

Branching fractions $\br[\hbn\to\etm\ga]$ are determined from
$N_{nP\to mS}^{\rm total}/(N_{nP}^{\rm total}\;\epsilon$), where
$N_{nP\to mS}^{\rm total}=N_{nP\to mS}+N_{nP\to mS}^{\rm anti-cut}$
and $N_{nP}^{\rm total}=N_{nP}+N_{nP}^{\rm anti-cut}$,
$N_{nP\to mS}^{\rm anti-cut}$ [$N_{nP}^{\rm anti-cut}$] is the number
of the $\hbn\to\etm\ga$ transitions [$\hbn$] that are rejected by the
$R_2$ and $\pi^0$ veto requirements [by the $R_2$ requirement],
$\epsilon$ is the reconstruction efficiency of the photon. In this
way, we do not rely on MC simulation for the determination of the
efficiency of the $R_2$ and $\pi^0$ veto requirements.
To determine $N_{nP}^{\rm anti-cut}$ and $N_{nP\to mS}^{\rm
  anti-cut}$, we repeat the analysis for the events
  rejected by the corresponding requirements. In the $\dmmppg$
[$\mmpp$] fits, we fix the $\etm$ mass and width [the $\hbn$ mass].
We estimate the $N_{nP}^{\rm anti-cut}$ systematic uncertainty by
varying the polynomial order and fit range in the $\mmpp$ fit and by
varying the $\hbn$ mass within its uncertainty. 
We find
$N_{1P}^{\rm total}=(84.2\pm4.4\,^{+2.1}_{-1.3})\times10^3$ and 
$N_{2P}^{\rm total}=(98.5\pm8.1\,^{+5.5}_{-6.3})\times10^3$.
For $N^{\rm total}$, we add the $N$ and $N^{\rm anti-cut}$
uncertainties in quadrature.
The contributions from the various sources of systematic uncertainty
to the $N_{nP\to mS}^{\rm total}$ are given in
Table~\ref{tab:syst_etn_yield}.
The $\hbn$ signal shape uncertainty is not considered for the yields,
since it cancels in the ratio
$N_{nP\to mS}^{\rm total}/N_{nP}^{\rm total}$.
In combining the $N_{nP\to mS}$ and $N_{nP\to mS}^{\rm anti-cut}$
uncertainties, we take into account correlations of the calibration
and $\etn$ line shape sources.
We find $N_{1P\to1S}^{\rm total}=(30.9\pm3.2\,^{+3.4}_{-1.8})\times10^3$,
$N_{2P\to1S}^{\rm total}=(16.1\pm2.4\,^{+2.0}_{-2.2})\times10^3$
and
$N_{2P\to2S}^{\rm total}=(35.6\pm7.3\,^{+4.5}_{-5.4})\times10^3$.
\begin{table}[htb]
\caption{ Systematic uncertainties in the total yields of the
  $\hb\to\et\gamma$, $\hbp\to\et\gamma$ and $\hbp\to\ett\gamma$
  transitions (in $10^3$).}
\label{tab:syst_etn_yield}
\renewcommand{\arraystretch}{1.4}
\begin{ruledtabular}
\begin{tabular}{l|ccc}
& $N_{1P\to1S}^{\rm total}$ & $N_{2P\to1S}^{\rm total}$ & $N_{2P\to2S}^{\rm total}$ \\
\hline
$\mmpp$ fits         & $^{+0.2}_{-1.1}$ & $^{+1.6}_{-1.8}$ & $^{+0.0}_{-1.3}$ \\
$\dmmppg$ binning    & $^{+1.0}_{-0.3}$ & $^{+0.5}_{-0.1}$ & $^{+1.5}_{-3.7}$ \\
$\dmmppg$ fit        & $^{+3.2}_{-1.2}$ & $^{+1.0}_{-1.3}$ & $^{+2.5}_{-2.1}$ \\
Calibration          & $^{+0.3}_{-0.4}$ & $\pm0.0$        & $\pm1.3$ \\
$\etn$ line shape    & $^{+0.2}_{-0.0}$ & $^{+0.1}_{-0.0}$ & $^{+0.3}_{-0.0}$ \\
$\etn$ mass \& width & $^{+0.7}_{-0.5}$ & $\pm0.3$        & $^{+3.1}_{-2.8}$ \\
\hline  
Total                & $^{+3.4}_{-1.8}$ & $^{+2.0}_{-2.2}$ & $^{+4.5}_{-5.4}$
\end{tabular}
\end{ruledtabular}
\end{table}

The efficiency is determined using phase-space MC simulated events
that we weight according to the expectation
$1-\cos^2\theta_{\ga}+2\cos\theta_{\ga}
\cos\theta_{\pip}\cos\theta_{\pip\ga}$~\cite{Voloshin_priv}, where
$\theta_{\ga}$ ($\theta_{\pip}$) is the angle between the beam axis
and $\ga$ ($\pip$) momentum and $\theta_{\pip\ga}$ is the angle
between $\ga$ and $\pip$ momenta, with all momenta measured in the
c.m.\ frame.
Efficiencies to reconstruct the photon after the $\pp$ pair is already
reconstructed are 74.6\% for $\hb\to\et\ga$, 73.4\% for
$\hbp\to\et\ga$ and 76.1\% for $\hbp\to\ett\gamma$.
The $\et$ [$\ett$] signal function is normalized to unity in the mass
window $9.3\,\gevm-9.5\,\gevm$ [$9.9\,\gevm-10.1\,\gevm$]; thus, the
yields and efficiencies correspond to this mass window.
The efficiencies have relative uncertainties of 2\% due to possible
differences between the data and MC simulation, and $^{+0.7}_{-0.8}$\%
[$^{+0.5}_{-0.6}$\%] due to the uncertainty in the $\et$ [$\ett$]
width.
We find
$\br[\hb\to\et\gamma]=(49.2\pm5.7\,^{+5.6}_{-3.3})\%$,
$\br[\hbp\to\et\gamma]=(22.3\pm3.8\,^{+3.1}_{-3.3})\%$ and
$\br[\hbp\to\ett\ga]=(47.5\pm10.5\,^{+6.8}_{-7.7})\%$.
These branching fractions are a factor of 1.2 to 2.5 higher than
theoretical expectations~\cite{godros}.

In summary, we report the first evidence for the $\ett$ using the
$\hbp\to\ett\ga$ transition, with a significance, including
systematics, of $4.2\,\sigma$, and the first observation of the
$\hb\to\et\ga$ and $\hbp\to\et\ga$ transitions. The mass and width
parameters of the $\et$ and $\ett$ are measured to be
$m_{\et}=(9402.4\pm1.5\pm1.8)\,\mevm$,
$m_{\ett}=(9999.0\pm3.5\,^{+2.8}_{-1.9})\,\mevm$ and
$\Gamma_{\et}=(10.8\,^{+4.0}_{-3.7}\,^{+4.5}_{-2.0})\,\mev$.
The $m_{\ett}$ and $\Gamma_{\et}$ are first measurements; the
$m_{\et}$ measurement is more precise than the current world average
and is $(11.4\pm3.6)\,\mevm$ above the central value~\cite{PDG}.
The hyperfine splittings, $\Delta M_{\rm HF}(1S)=(57.9\pm2.3)\,\mevm$,
$\Delta M_{\rm HF}(2S)=(24.3^{+4.0}_{-4.5})\,\mevm$ and their ratio
$0.420^{+0.071}_{-0.079}\,$, are in agreement with theoretical 
calculations~\cite{Lattice,Lattice2}.
We update the $\hb$ and $\hbp$ mass measurements
$m_{\hb}=(9899.1\pm0.4\pm1.0)\,\mevm$,
$m_{\hbp}=(10259.8\pm0.5\pm1.1)\,\mevm$, and $1P$ and $2P$ hyperfine
splittings $\Delta M_{\rm HF}(1P)=(+0.8\pm1.1)\,\mevm$, $\Delta M_{\rm
  HF}(2P)=(+0.5\pm1.2)\,\mevm$.
These results supersede those in Ref.~\cite{Belle_hb}.

We are grateful to A.~M.~Badalian, Yu.~S.~Kalashnikova and
M.~B.~Voloshin for useful discussions.
We thank the KEKB group for excellent operation of the
accelerator; the KEK cryogenics group for efficient solenoid
operations; and the KEK computer group, the NII, and 
PNNL/EMSL for valuable computing and SINET4 network support.  
We acknowledge support from MEXT, JSPS and Nagoya's TLPRC (Japan);
ARC and DIISR (Australia); NSFC (China); MSMT (Czechia);
DST (India); INFN (Italy); MEST, NRF, GSDC of KISTI, and WCU (Korea); 
MNiSW (Poland); MES and RFAAE (Russia); ARRS (Slovenia); 
SNSF (Switzerland); NSC and MOE (Taiwan); and DOE and NSF (USA).


\begin{thebibliography}{99}
\bibitem{mmpp_fit_details} Comparing to Ref.~\cite{Belle_hb}, the
  $\zb$ requirement changed the shape of the combinatorial
  background. Therefore, the polynomial order was modified to obtain
  the highest C.L. The $\hb$ and $\hbp$ regions are described,
  respectively, by a 3rd order and 8th order Chebyshev polynomial
  instead of a 5th and 7th order. In addition, the fit interval is
  reduced by decreasing the end point from $10.45\,\gevm$ to
  $10.40\,\gevm$.
\end{thebibliography}
\end{document}